\documentclass[12pt,preprint]{aastex}
\usepackage{epsfig, amsmath, amssymb}

\def\be{\begin{equation}}
\def\ee{\end{equation}}
\def\ba{\begin{eqnarray}}
\def\ea{\end{eqnarray}}
\def\go{\gtrsim}
\def\lo{\lesssim}
\def\cJ{{\cal J}}

\newcommand{\dnachd}[2]{\frac{\partial #1}{\partial #2}}
\newcommand{\hatl}{\hat{\bf l}}
\newcommand{\ve}[1]{{\bf #1}}

\begin{document}

\title{Warping and Precession of Accretion Disks Around Magnetic Stars: 
Nonlinear Evolution}
\author{Harald P. Pfeiffer\altaffilmark{1} and Dong Lai}
\affil{Center for Radiophysics and Space Research, Cornell University,
Ithaca, NY 14853\\
Email: harald,dong@astro.cornell.edu}
\altaffiltext{1}{Current address: Theoretical Astrophysics 130-33, California Institute of Technology, Pasadena, California 91125}

\begin{abstract}
The inner region of the accretion disk around a magnetized star 
(T Tauri star, white dwarf or neutron star) is
subjected to magnetic torques that induce warping and precession of the
disk. These torques arise from the interaction between the stellar field and
the induced electric currents in the disk. We carry out numerical 
simulations of the nonlinear evolution of warped, viscous 
accretion disks driven by the magnetic torques. We show that 
the disk can develop into a highly warped steady state 
in which the disk attains a fixed (warped) shape and precesses rigidly.
The warp is most pronounced at the disk inner radius (near the magnetosphere
boundary). As the system parameters (such as accretion rate) change,
the disk can switch between a completely flat state (warping stable) and a 
highly warped state. The precession of warped disks may be 
responsible for a variety of quasi-periodic oscillations or radiation
flux variabilities observed in many different systems, including 
young stellar objects and X-ray binaries. 
\end{abstract}

\keywords{accretion, accretion disks -- stars: neutron --
stars: magnetic fields -- stars: pre-main-sequence -- binaries: general}

\section{Introduction}

Warped accretion disks are believed to exist in a variety of astrophysical
systems, including X-ray binaries, young stellar objects (YSOs) and
active galactic nuclei. For example, the well-known 164 day precession of 
the jet in SS 433 and the long-term variabilities observed in many X-ray 
binaries (e.g. Her X-1, LMC X-4; see Priedhorsky \& Holt 1987; 
Scott et al.~2000; Ogilvie \& Dubus 2001) have been 
explained by the precession of tilted accretion disks. The changes
in the flow directions of several YSO jets (e.g. Bate et al.~2000) 
and the photometric variabilities of some T Tauri stars (e.g., AA Tauri;
see Bouvier et al.~1999; Carpenter et al.~2001) have also been associated 
with warped, precessing disks. 

Several mechanisms for exciting/maintaining warps in accretion disks
have been proposed in recent years. Schandl \& Meyer (1994) considered
a warping instability due to irradiation-driven disk wind. 
Pringle (1996) showed that even without wind loss, radiation pressure itself
can induce warping in the outer region of the disk
(see also Maloney, Begelman \& Nowak 1998; Wijers \& Pringle 1999;
Ogilvie \& Dubus 2001). Quillen (2001) showed that a wind 
passing over the disk surface may induce warping via 
Kelvin-Helmholtz instability. Finally, in the case of disk accretion 
onto magnetic stars (e.g., neutron stars, white dwarfs and T Tauri stars), 
the stellar magnetic field can induce disk warping and precession 
(Lai 1999; see also Terquem \& Papaloizou 2000). In this paper we are
concerned with these magnetic effects on accretion disks.

The study of disk accretion onto magnetic stars has a long history
(e.g., Pringle \& Rees 1972; Ghosh \& Lamb 1979; Anzer \& B\"orner
1980,~1983; Lipunov \& Shakura 1980; Wang 1987,~1995; Aly \& Kuijpers
1990; Spruit \& Taam 1993; Shu et al.~1994; van Ballegooijen 1994;
Lovelace et al.~1995,1999; Campbell 1997; Lai 1998; see also numerical
simulations by Hayashi et al.~1996; Miller \& Stone 1997; Goodson et
al.~1997; Fendt \& Elstner 2000; Romanova et al.~2003). Most previous
studies have, for simplicity, adopted the idealized geometry in which
the magnetic axis, the spin axis and the disk angular momentum are
aligned. However, it was shown in Lai (1999) that under quite general
conditions, the stellar magnetic field can induce warping in the inner
disk and make the disk precess around the spin axis (see \S 2; see
also Aly 1980; Lipunov \& Shakura 1980; Terquem \& Papaloizou 2000).
Such magnetically driven warping and precession open up new
possibilities for the dynamical behaviors of disk accretion onto
magnetic stars.  Shirakawa \& Lai (2002a) studied these effects in
weakly magnetized accreting neutron stars and showed that the magnetic
warping/precession effects may explain several observed features of
low-frequency quasi-periodic oscillations in low-mass X-ray
binaries. Shirakawa \& Lai (2002b) also studied the linear, global
warping/precession modes of inner disks of highly magnetized ($B\sim
10^{12}$~G) neutron stars (NSs), as in accreting X-ray pulsars, and
suggested that magnetically driven disk warping and precession can
explain the mHz variabilities observed in X-ray pulsars.

The studies by Shirakawa \& Lai (2002a,b) were restricted to linear
warping. Such linear analysis is only the first step toward understanding the
observational manifestations of the magnetic warping/precession effects.
However, as we show in this paper, under many circumstances, 
the disk warping becomes nonlinear, and the dynamics of nonlinear warped 
disks is much richer. This is the subject of this paper.
After briefly reviewing the basic physics of magnetically driven disk
warping and precession (\S2), we present the dynamical equations for
the warped disk (\S3) and our numerical results (\S 4), and discuss 
their astrophysical implications (\S5). The appendix contains more details
of our numerical method.

\section{Magnetically Driven Disk Warping/Precession}

For completeness, we briefly review the basic physics of
magnetically driven warping/precession and give the formulae
to be used in later sections.

The inner region of the accretion disk onto a rotating magnetized central star
is subjected to magnetic torques that induce warping and precession of the
disk (Lai 1999). These magnetic torques result from the interactions between
the accretion disk and the stellar magnetic field.
Depending on how the disk responds to the stellar field,
two different kinds of torque arise:
(i) If the vertical stellar magnetic field $B_z$ penetrates the disk,
it gets twisted by the disk rotation
to produce an azimuthal field $\Delta B_\phi=\mp\zeta B_z$ that has different
signs above and below the disk ($\zeta$ is the azimuthal pitch of the field
line and depends on the dissipation in the disk), and a radial surface current
$K_r$ results. The interaction between $K_r$ and the stellar $B_\phi$ gives
rise to a vertical force. While the mean force (averaging over the azimuthal
direction) is zero, the uneven distribution of the force induces a net
{\it warping torque} which tends to misalign the angular momentum of the disk
with the stellar spin axis.
(ii) If the disk does not allow the vertical stellar field
(e.g., the rapidly varying component of $B_z$ due to stellar rotation)
to penetrate, an azimuthal screening current $K_\phi$ will be induced on the
disk. This $K_\phi$ interacts with the radial magnetic field $B_r$
and produces a vertical force. The resulting {\it precessional torque}
tends to drive the disk into retrograde precession around the stellar spin
axis.

In general, both the magnetic warping torque and the precessional torque are
present. For small disk tilt angle $\beta$ (the angle between the disk normal
and the spin axis), the precession angular frequency and warping rate
at radius $r$ are given by (Lai 1999)
\ba
&&\Omega_p (r)=\frac{\mu^2}{\pi^2 r^7\Omega(r)\Sigma(r) D(r)}F(\theta),
\label{eqn:Omega_p}\\
&&\Gamma_w (r)=\frac{\zeta\mu^2}{4\pi r^7\Omega(r)\Sigma(r)}\cos^2\theta,
\label{eqn:Gamma_w}
\ea
where $\mu$ is the stellar magnetic dipole moment, $\theta$ is
the angle between the magnetic dipole axis and the spin axis,
$\Omega(r)$ is the orbital angular frequency, and $\Sigma(r)$ is the surface
density of the disk. The dimensionless function $D(r)$ is given by
\begin{equation}
D(r)={\rm max}~\left(\sqrt{r^2/r^2_{in}-1}, \sqrt{2H(r)/r_{in}}\right)
\label{eqn:D(r)}.
\end{equation}
Here $H(r)$ is the half-thickness of the disk, and $r_{in}$ is the 
inner disk radius, given by the magnetosphere radius
\begin{equation}
r_{in}\equiv\eta\left(\frac{\mu^4}{GM\dot M^2}\right)^{1/7},
\label{eqn:r_m}
\end{equation}
where $M$ is the stellar mass, ${\dot M}$ is
the mass accretion rate, and $\eta\sim0.5-1$.
The function $F(\theta)$ depends on the dielectric property of the disk. 
We can write
\begin{equation}
F(\theta)=2f\cos^2\theta-\sin^2\theta,
\end{equation}
so that $F(\theta)=-\sin^2\theta$ if only the spin-variable vertical field is
screened out by the disk ($f=0$), and $F(\theta)=3\cos^2\theta-1$ if all 
vertical field is screened out ($f=1$). In reality, $f$ lies between 0 and~1.
For concreteness, we shall set $F(\theta)=-\sin^2\theta$ in the following.

\section{Dynamical Equations for Warped Disks}

In a viscous accretion disk (with the dimensionless viscosity parameter
$\alpha\gtrsim H/r$), the dynamics of the disk warp is dominated
by viscous stress (rather than bending waves; see
Papaloizou \& Lin 1995; Terquem 1998), and can be studied using
the formalism of Papaloizou \& Pringle (1983) (see also Pringle~1992;
Ogilvie~1999; Ogilvie \& Dubus~2001). We model the disk as a collection of
rings which interact with each other via viscous stresses.
Each ring at radius $r$ has the unit normal vector ${\bf\hat l}(r,t)$.
In the Cartesian coordinates, with the $z$-axis along the stellar spin
${\hat\omega}$, we can write
\begin{equation}\label{eq:hatl}
\hat{\bf l}=\begin{pmatrix}
	\sin\beta\cos\gamma \\
	\sin\beta\sin\gamma\\
	\cos\beta
\end{pmatrix},
\end{equation}
with $\beta(r,t)$ the tilt angle and $\gamma(r,t)$ the twist angle.
The relevant equations are mass conservation,
\begin{equation}\label{eq:Evolution-Sigma}
\dnachd{\Sigma}{t}+\frac{1}{r}\dnachd{}{r}\left(r\Sigma V_r\right)=0,
\end{equation}
and the evolution equation for $\hat{\bf l}$,
\begin{equation}\label{eq:Evolution-hatl}
\dnachd{\hat{\bf l}}{t} +\left[V_r-\frac{\nu_1\dnachd{\Omega}{r}}{\Omega}
-\frac{1}{2}\nu_2\frac{\dnachd{}{r}(\Sigma r^3\Omega)}{\Sigma r^3\Omega}
\right]\dnachd{\hat{\bf l}}{r}
=\dnachd{}{r}{\left(\frac{1}{2}\nu_2\dnachd{\hat{\bf l}}{r}\right)
+\frac{1}{2}\nu_2\left|\dnachd{\hat{\bf l}}{r}\right|^2\hat{\bf l}}
+\frac{\bf N}{\Sigma r^2\Omega}.
\end{equation}
Here $\nu_1$ is the usual disk velocity (measuring the $r$--$\phi$
stress), $\nu_2$ denotes the viscosity in the $r$--$z$ stress which is
associated with reducing the disk tilt and $\ve N$ represents
external torques acting on the disk.  The radial velocity was derived
by Pringle (1992),
\begin{equation}\label{eq:Vr}
V_r=\frac{\dnachd{}{r}\left(\nu_1\Sigma r^3\dnachd{\Omega}{r}\right)}
{r\Sigma\dnachd{}{r}\left(r^2\Omega\right)}
-\frac{\nu_2}{2}\frac{r^2\Omega}{\dnachd{}{r}(r^2\Omega)}\left|\dnachd{\hatl}{r}\right|^2.
\end{equation}
The nonlinear warp equation (\ref{eq:Evolution-hatl}) is based on a
phenomenological description of the viscosities $\nu_1$ and $\nu_2$
and is formally valid only in the linear regime when the local disk
warp $\partial\hatl/\partial\ln r$ is small (Ogilvie 1999).  A more
rigorous treatment by Ogilvie (1999) removes this assumption and finds
that $\nu_1$ and $\nu_2$ effectively depend on the warp amplitude
$\partial\hatl/ \partial\ln r$.  Ogilvie (1999) also finds an
additional term causing dispersive propagation and precession.
However, it turns out that this latter term is not very important for
Keplerian disks.  For the current investigation into the qualitative
features of the magnetic warping and precession instability we adopt
the simpler equations given by Pringle (1992).

The torque ${\bf N}$ in Eq.~(\ref{eq:Evolution-hatl}) was derived in
Lai (1999) as
\begin{equation}\label{eq:unscaled-N}
\frac{\bf N}{\Sigma r^2\Omega}
=\Omega_p \cos\beta\;\hat{\omega}\times\hat{\bf l}
-\Gamma_w\cos\beta\,\big[\hat\omega-(\hat\omega\cdot\hatl)\hatl\,\big]
\end{equation}
where $\Omega_p$ and $\Gamma_w$ are given by Eqs.~(\ref{eqn:Omega_p})
and (\ref{eqn:Gamma_w}).  The first term of Eq.~(\ref{eq:unscaled-N})
causes retrograde precession of the tilted disk (since $\Omega_p<0$
due to the choice $F(\theta)=-\sin^2\theta$), whereas the second term
tries to warp the disk.  

The magnetic torque formulae [eqs.~(\ref{eqn:Omega_p}) and
(\ref{eqn:Gamma_w})] contain uncertain parameters (e.g., $\zeta$,
which parametrizes the amount of azimuthal twist of the magnetic field
threading the disk); this is inevitable given the complicated nature
of magnetic field -- disk interaction (see Lai 1999 and references
therein). Also, while the expression for the warping torque
[eq.~(\ref{eqn:Gamma_w})] is formally valid for large disk warps, the
expression for the precession torque was derived under the assumption
that the disk is locally flat [eq.~(\ref{eqn:Omega_p}) is strictly
valid only for a completely flat disk; see Aly 1980]; when this
assumption breaks down (i.e., when $\partial\hatl/\partial\ln r$ is
large), we expect a similar torque expression to hold, but with
modified numerical factors (e.g. the function $D(r)$ in
eq.~(\ref{eqn:Omega_p}) will be different).  When an almost flat disk
becomes unstable to the warping instability, and also in the deeply
nonlinear warp regime with disk tilted to large radii (see \S 4), the
condition $|\partial\hatl/\partial\ln r|\lo 1$ is well satisfied.
Thus we believe that our simplifying assumptions capture the
qualitative behavior of accretion disks subject to the magnetic
torques.

\subsection{Boundary Conditions}

We need boundary conditions for $\hatl$ and for the surface density
$\Sigma$.  The torque~(\ref{eq:unscaled-N}) decays rapidly with
radius, so that we expect the outer disk to be undisturbed.  This is
confirmed by the linear stability analysis of Shirakawa \& Lai
(2002a,b).  We therefore adopt simple Dirichlet boundary conditions at
the outer edge of the disk,
\begin{equation}\label{eq:BC-out}
\hatl(r_{out})\equiv \hatl_{out}=\mbox{const.}
\end{equation}
Throughout this paper we choose $\hatl_{out}=\hat\omega$, such that
the outer disk is perpendicular to the rotation axis of the star.  At
the inner edge of the disk, we use a zero torque boundary condition,
\begin{equation}\label{eq:BC-in}
\dnachd{\hatl}{r}(r_{in})=0.
\end{equation}
The boundary condition on $\Sigma$ is motivated by the flat Keplerian
disk, which has\footnote{This expression for $\Sigma_{flat}$ applies
to a nonmagnetic disk.  When magnetic fields thread the disk, the
functional form of $\cJ(r)$ is modified in a model-dependent way (see
Lai 1999, appendix A for examples).}
\begin{equation}\label{eq:Sigma-flat}
\Sigma_{flat}=\frac{\dot M}{3\pi\nu_1}\cJ,
\end{equation}
with $\cJ=1-(1-\cJ_{in})\left(r/r_{in}\right)^{-1/2}$. We choose
Dirichlet boundary conditions on the surface density,
\begin{equation}
\Sigma(r_{out})=\frac{\dot M}{3\pi\nu_1}\cJ_{out},\qquad
\Sigma(r_{in})=\frac{\dot M}{3\pi\nu_1}\cJ_{in},
\end{equation}
where $\cJ_{out}=1-(1-\cJ_{in})(r_{out}/r_{in})^{-1/2}$.
We take $\cJ_{in}$ to be free, parametrizing the physics at the inner
edge of the disk.

\subsection{Dimensionless Equations}

The radius is measured in multiples of the radius of the inner edge,
$x=r/r_{in}$, and we introduce a dimensionless surface density
$\sigma=\Sigma/\Sigma_0$.  We assume viscosities of the form
\begin{align}\label{eq:dimensionless-nu}
\nu_1=\nu_{10}x^{\beta_1}\sigma^{\sigma_1},\quad
\nu_2=\nu_{20}x^{\beta_2}\sigma^{\sigma_2}.
\end{align}
In light of Eq.~(\ref{eq:Sigma-flat}), we take $\Sigma_0=\dot
M/(3\pi\nu_{10})$.  Finally, we measure time in units of the viscous
time-scale at the inner edge, $t=t_0\tau$, $t_0=r_{in}^2/\nu_{10},$
and define $\eta_0=\nu_{20}/\nu_{10}.$ With these assumptions,
Eqs.~(\ref{eq:Evolution-Sigma})--(\ref{eq:Vr}) simplify to
\begin{align}
\label{eq:sigma2}
\dnachd{\sigma}{\tau}
+\frac{1}{x}\dnachd{}{x}\left[
      \frac{\left(\sigma^{\sigma_1+1}x^{\beta_1+3}\Omega'\right)'}
	   {(x^2\Omega)'}
      -\frac{\eta_0}{2}\sigma^{\sigma_2+1}
          \frac{x^{\beta_2+3}\Omega}{(x^2\Omega)'}\left|\hatl'\right|^2
  \right]
&=0,\\
\nonumber
\dnachd{\hatl}{\tau}
-\frac{\eta_0}{2}\sigma^{\sigma_2}x^{\beta_2}
  \left(\hatl''+\left|\hatl'\right|^2\hatl\right)
+\Bigg\{\frac{\left(\sigma^{\sigma_1+1}x^{\beta_1+3}\Omega'\right)'}
            {\sigma x(x^2\Omega)'}
        -\sigma^{\sigma_1} x^{\beta_1}\frac{\Omega'}{\Omega}\qquad\qquad&\\
        -\frac{\eta_0}{2}\left[
	    \frac{\left(\sigma^{\sigma_2+1}x^{\beta_2+3}\Omega\right)'}
                 {\sigma x^3\Omega}
            +\sigma^{\sigma_2}\frac{x^{\beta_2+2}\Omega}{(x^2\Omega)'}
	     \left|\hatl'\right|^2
         \right]
\Bigg\}\,\hatl'
&=\ve n,
\label{eq:lhat2}
\end{align}
where a prime denotes $\partial/\partial x$.  The dimensionless
torque is given by
\begin{equation}\label{eq:n}
\ve n=\frac{t_0\ve N}{r^2\Sigma\Omega}
=\frac{\hatl\cdot\hat\omega}{\sigma x^{5.5}}\left[-\frac{\bar\Omega_p}{D(x)}\;\hat\omega\times\hatl
-\bar\Gamma_w\left(\hat\omega-(\hat\omega\cdot\hatl)\hatl\right)\right],
\end{equation}
where
 \ba 
&&\bar\Omega_p
=\frac{3}{\pi}\frac{\sin^2\theta}{\eta^{3.5}}
=5.4\left(\frac{\eta}{0.5}\right)^{-3.5}
  \left(\frac{\sin^2\theta}{0.5}\right),\\
&&\bar\Gamma_w=\frac{3}{4}\eta^{-3.5}\zeta\cos^2\theta
=21\left(\frac{\eta}{0.5}\right)^{-3.5}\left(\frac{\zeta}{5}\right)
\left(\frac{\cos^2\theta}{0.5}\right)  \ea 
 are dimensionless constants
independent of $r$ and $t$, parametrizing the problem.
We will first set $D(x)=1$ to make contact to the linear stability
analysis in Shirakawa \& Lai (2002b).  Later in this paper, we will
use Eq.~(\ref{eqn:D(r)}) and set for concreteness
$H(r)/r_{in}=0.02$ close to the inner edge, so that
\begin{equation}
\label{eq:D} D(x)=\max\left(\sqrt{x^2-1}\;,\;\; 0.2\right).
\end{equation}
(The specific choice of $D(x)$ influences our qualitative results only
marginally, as the radial dependence of the precession torque
in Eq.~(\ref{eqn:Omega_p}) is dominated by the $r^{-7}$ piece.)

While eqs.~(\ref{eq:sigma2}) and (\ref{eq:lhat2}) can be applied 
for general viscosity law [eq.~(\ref{eq:dimensionless-nu})], in our
calculations we shall restrict to the cases 
where $\nu_2/\nu_2=\nu_{20}/\nu_{10}=\eta_0$. The ratio $\eta_0$
measures the disk's resistance to warping; we will consider
different values of $\eta_0$, ranging from $\eta_0=1$ to $\eta_0\gg 1$
\footnote{For a Keplerian disk, assuming isotropic 
viscous stress parametrized by the $\alpha$-ansatz, 
the ratio $\nu_2/\nu_1$ is of order $1/(2\alpha^2)$ for linear warps
(Papaloizou \& Pringle 1983; Ogilvie 1999). The large $\nu_2/\nu_1$
(for $\alpha\ll 1$) is due to the fact that horizontal motions induced in 
the disk by the warp are resonantly driven. Such resonant behavior 
tends to be diminished in the nonlinear regime and also when
deviation from Keplerian rotation is present (Ogilvie 1999).}.

Finally, we give the boundary conditions in dimensionless variables:
\begin{align}\label{eq:BC-lhat}
\dnachd{\hatl}{x}(x_{in})&=0,&\hatl (x_{out})&=\hatl_{out}\\
\label{eq:BC-sigma}
\sigma(x_{in})&=\sigma_{flat}(x_{in}),&
\sigma(x_{out})&=\sigma_{flat}(x_{out}),
\end{align}
where, by definition, $x_{in}=1$, and $\sigma_{flat}\equiv \Sigma_{flat}/\Sigma_0$ follows from (\ref{eq:Sigma-flat}) and
(\ref{eq:dimensionless-nu}):
\begin{equation}
\sigma_{flat}=\left[\left(1-(1-\cJ_{in})x^{-1/2}\right)x^{-\beta_1}\right]^{1/(1+\sigma_1)}.
\end{equation}

\section{Numerical Results}

Equations~(\ref{eq:sigma2}) and~(\ref{eq:lhat2}) with boundary
conditions given by Eqs.~(\ref{eq:BC-lhat}) and~(\ref{eq:BC-sigma}) are
solved with a Crank-Nicholson scheme as described in the appendix.

\subsection{Evolution Into the Nonlinear Regime}
\label{sec:LinearEvolution}

We first consider an almost flat, unstable disk which we follow
through the linear growth of the warping mode into the nonlinear
regime. The particular parameters we use are $\bar\Omega_p=10$,
$\bar\Gamma_w=10$, $\cJ_{in}=1$, $\eta_0=1$, $\beta_1=\beta_2=0.6$,
$\sigma_1=\sigma_2=0$ and set $D(x)=1$.  The linear stability analysis
for these parameters was presented in Shirakawa \& Lai (2002b), where
the disk was found to be unstable to the magnetic warping instability.
As initial conditions for the evolution we take
$\left.\sigma(x)\right|_{\tau=0}=\sigma_{flat}(x)$, and
perturb~$\hatl$ by $\sim 10^{-5}$ degrees away from $\hat\omega$.

\begin{figure}
\epsscale{0.6}
\plotone{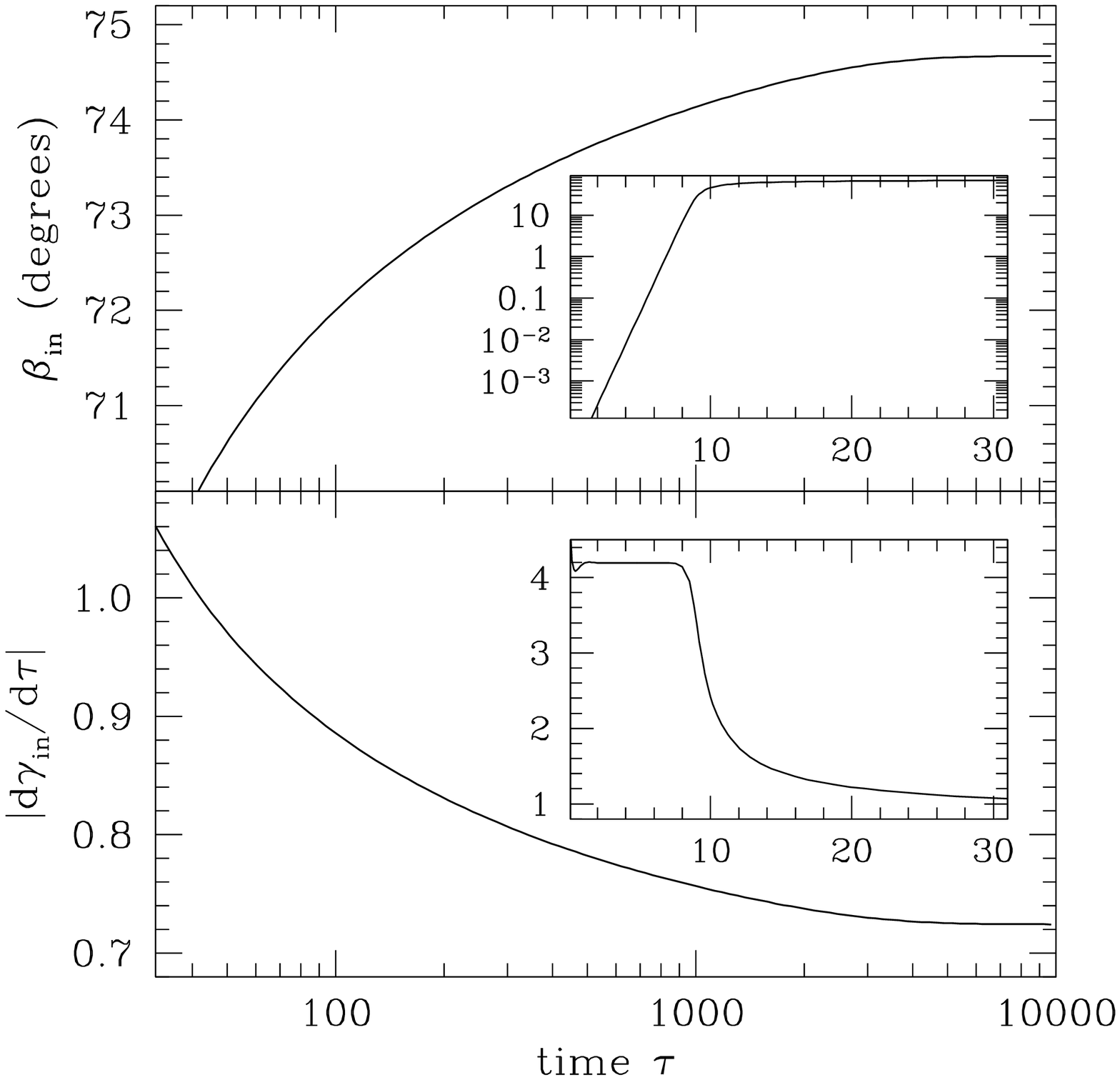}
\caption{\label{fig:early_times}Tilt angle $\beta_{in}$ at the inner
edge and precession frequency at the inner edge during evolution into
the nonlinear regime (parameters $\bar\Omega_p=\bar\Gamma_w=10$, $\cJ_{in}=1$,
$\eta_0=1$, $\beta_1=\beta_2=0.6$, $\sigma_1=\sigma_2=0$.)  The
inserts show early times.}
\end{figure}

Fig.~\ref{fig:early_times} presents the tilt-angle at the inner edge,
$\beta_{in}=\beta(x_{in})$ and the precession frequency at the inner
edge, $d\gamma_{in}/d\tau$ as a function of time $\tau$.  The disk is
indeed unstable and a growing mode emerges.  At early times, the
tilt-angle grows exponentially with e-folding time of
$T_{\mbox{\footnotesize e-fold}}=(d\log\beta_{in}/d\tau)^{-1}=0.59$ and 
precession frequency $d\gamma_{in}/d\tau=-4.22$.  These numbers agree
very well with the linear stability analysis of Shirakawa \& Lai
(2002b).  At late times the growth slows down, and the tilt-angle
saturates around $\beta(r_{in})\approx 75^\circ$.  Eventually, the
disk settles into a solution where the whole disk precesses with
uniform precession frequency, i.e. the tilt-angle profile $\beta(r)$
is independent of time, whereas the twist angle increases linearly in
time, $\gamma(r,\tau)=\gamma_{0}(r)+\omega_p\tau$, with the precession
frequency $\omega_p\simeq -0.72$.

\begin{figure}
\epsscale{0.6}
\plotone{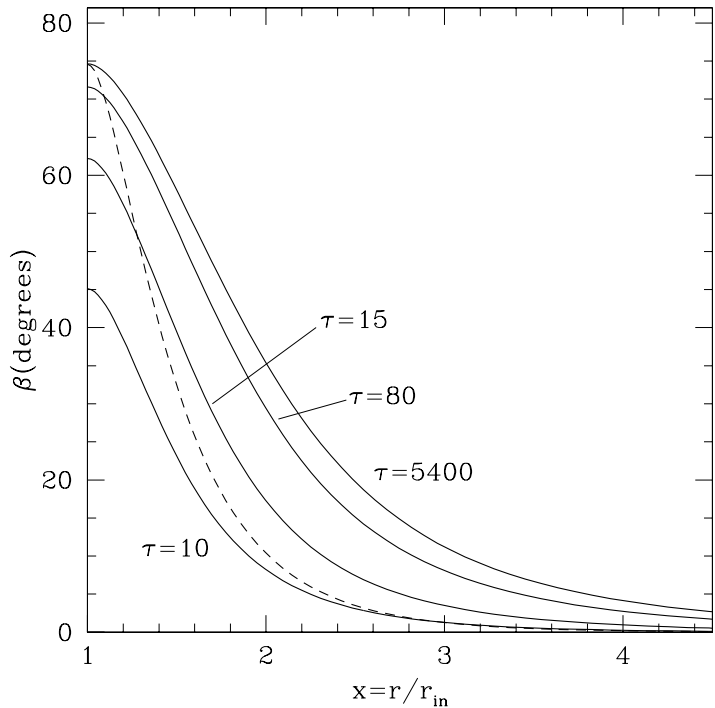}
\caption{\label{fig:beta_profiles}Tilt angle $\beta$ as a function of
radius at different times during the evolution presented in
Fig. \ref{fig:early_times}.  The solid lines illustrate the growth of
the warp; at $\tau=5400$, the disk has essentially reached its final
steady state.  The dashed line represents the shape during the linear
regime ($\tau=6$, rescaled to coincide with the steady state-solution
at the inner edge), which agrees well with the eigenvector
of the linear stability analysis of Shirakawa \& Lai (2002b).}
\end{figure}

Fig.~\ref{fig:beta_profiles} presents the tilt-angle $\beta(r)$ as a
function of radius at different times.  One sees clearly that after
the mode has saturated, the tilted region of the disk is larger than
during the linear phase.  The profiles of the tilt-angle provide an
explanation for the reduction of precession frequency as the disk
enters the non-linear regime: The precessional torque given by the
first term of Eq.~(\ref{eq:n}), is suppressed by the factor $\hat{\bf
l}\cdot\hat\omega=\cos\beta$. Furthermore, a more extended disk is
involved, reducing the precession rate further.  In the final
steady-state, the precession period is $P=2\pi/|\omega_p|=8.7$, while
the warp diffusion time $T_{\mbox{\footnotesize diff}}$ is of order
$|\partial\hatl/\partial r|^{-2}\nu_2^{-1}\sim r^2/\nu_2=
x^{1.4}/\nu_{20}$ (in dimensionless units).  The disk warp extents to
$x\lo 3$ (see Fig.~\ref{fig:beta_profiles}) so that
$T_{\mbox{\footnotesize diff}}\approx 4.7<P$.  Thus the different
parts of the warped disk can communicate with each other through
viscous diffusion on a timescale shorter than the precession time,
making rigid-body precession possible.  Also, both $P$ and
$T_{\mbox{\footnotesize diff}}$ are much larger than the e-folding
time $T_{\mbox{\footnotesize e-fold}}\approx 0.59$ of the growth of
linear perturbations.  The inequalities $T_{\mbox{\footnotesize
e-fold}}< T_{\mbox{\footnotesize diff}}< P$ hold for all evolutions
presented in this paper.

\begin{figure}
\epsscale{0.6}
\plotone{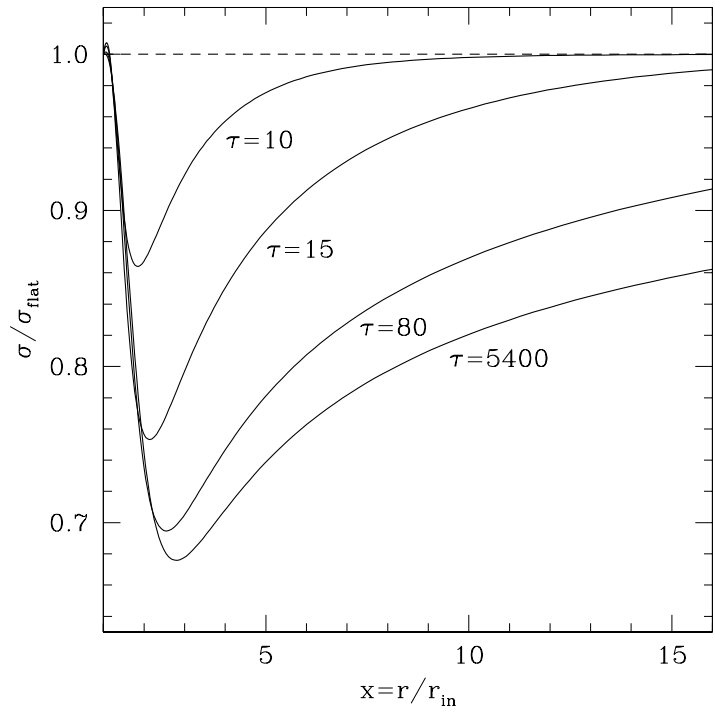}
\caption{\label{fig:sigma_profiles}Surface density relative to the
unperturbed, flat disk at different times during the evolution
presented in Fig. \ref{fig:early_times}. The dashed line represents
our initial condition, $\sigma/\sigma_{flat}\equiv 1$. }
\end{figure}

Fig.~\ref{fig:sigma_profiles} shows the surface density relative to
the flat disk, $\sigma/\sigma_{flat}$ at different times.  As the
evolution proceeds, the disk gets depleted.  The reason for this is
that as the disk warps, the second term in Eq.~(\ref{eq:Vr}) gives
rise to increased advection relative to the flat disk.  Although the
disk-tilt is confined to relatively small radii (e.g., $\beta<2^\circ$
for $x\gtrsim 5$; and $\beta$ decreases exponentially with larger
$x$), the surface density is depleted over a larger region with the
deviation from $\sigma_{flat}$ decreasing only as $x^{-1/2}$.  The
outer disk depletes on the viscous time-scale of the {\em outer} disk;
this explains the very slow relaxation to the steady-state in
Fig.~\ref{fig:early_times}. The fact that $\sigma$ changes at large
radii necessitates the use of a large computational domain, although
the disk is only warped in the inner region.  We typically place the
outer boundary at $x_{out}=1000$, which results in accuracy of about
$1\%$.

\subsection{Nonlinear Steady State of Warped Disk}

The growing mode is only a transient phenomenon --- one will most
likely observe only the nonlinear steady state solution. Therefore we
now study the properties of the steady-state solution as a function of
four of the parameters that characterize the system: $\bar\Gamma_w$,
$\bar\Omega_p$, $\cJ_{in}$ and $\eta_0$ (for definiteness, the other
parameters are fixed as in \S 4.1). 

\begin{figure}
\epsscale{0.6}
\plotone{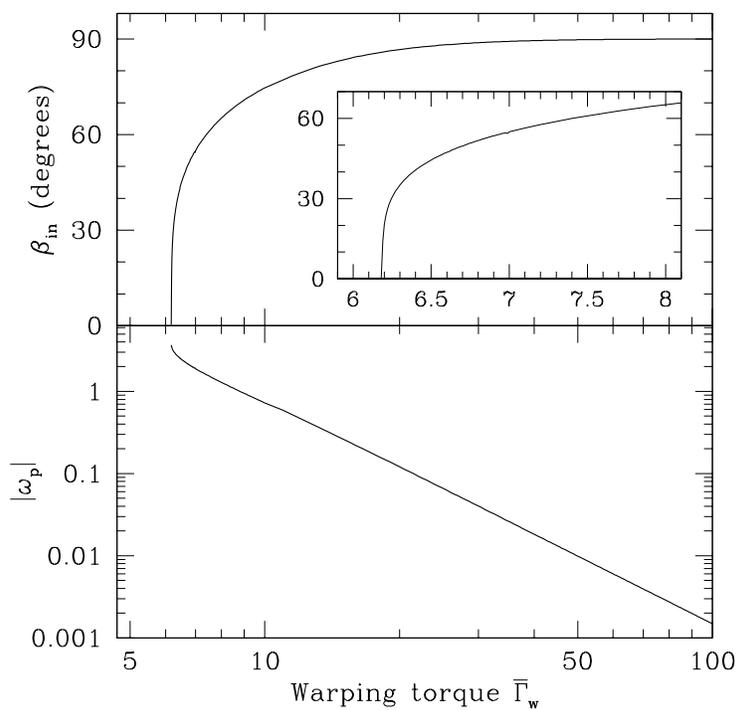}
\caption{\label{fig:vary_GammaW}The tilt-angle at the inner edge of
the disk, $\beta_{in}$, and the disk precession frequency $|\omega_p|$
as a function of warping torque parameter $\bar\Gamma_w$.  The other
parameters are fixed to $\bar\Omega_p=10, \cJ_{in}=1, \eta_0=1,
\beta_1=\beta_2=0.6, \sigma_1=\sigma_2=0$ and $D(x)=1$. }
\end{figure}

First we consider the effect of varying $\bar\Gamma_w$ (keeping
$\bar\Omega_p=10$, $\cJ_{in}=1$ and $\eta_0=1$ and $D(x)=1$ as in \S
4.1).  For different $\bar\Gamma_w$, we perform evolutions until
relaxation to steady-state, and then record the tilt angle
$\beta_{in}$ and the disk precession frequency
$\omega_p=d\gamma_{in}/d\tau$.  Figure \ref{fig:vary_GammaW} presents
the results of this computation.  For $\bar\Gamma_w\gtrsim 6.2$, the
disk is unstable to the magnetic warping instability (in agreement
with the linear stability analysis); with increasing torque
$\bar\Gamma_w$, the steady-state tilt-angle $\beta_{in}$ increases.
There is only a small window of torque parameters, for which the
steady-state disk is moderately warped (say, $0<\beta_{in}<
60^\circ$).  Hence, this computation suggests that, whenever the
magnetic warping instability operates, it is likely that the inner
disk will be significantly tilted.

\begin{figure}
\epsscale{0.6}
\plotone{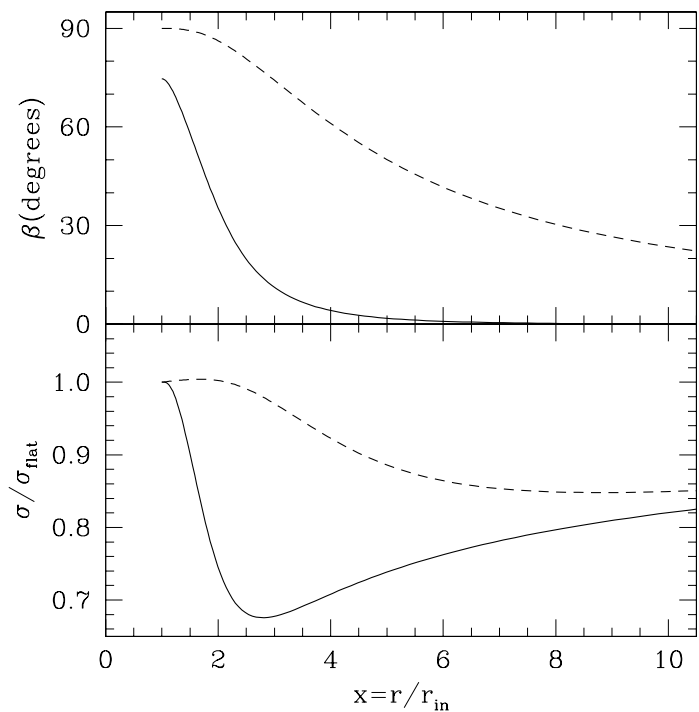}
\caption{\label{fig:compare_profiles}The
steady-state disk profiles for $\bar\Gamma_w=10$ (solid lines) and
$\bar\Gamma_w=100$ (dashed lines), with the other parameters the same as in
Fig.~\ref{fig:vary_GammaW}.  The top panel shows the tilt-angle
$\beta$ as a function of radius, the bottom panel the surface density.
Note that for $\bar\Gamma_w=100$, the disk is tilted out to larger
radii, but the surface density deviates less from that of the flat disk. }
\end{figure}

Figure \ref{fig:vary_GammaW} also shows the steady-state precession
frequency $\omega_p$ as a function of $\bar\Gamma_w$.  At the onset of
warping instability ($\bar\Gamma_w\simeq 6.2$), $\omega_p\approx
-3.7$. As $\bar\Gamma_w$ increases, $|\omega_p|$ drops very rapidly:
Doubling the warping torque decreases $|\omega_p|$ by roughly a factor
of ten. The reason is that for larger $\bar\Gamma_w$, the disk tilt is
larger (approaching $90^\circ$ very closely) and the warping region is
more extended (see Fig.~\ref{fig:compare_profiles}).  Note that the
surface density of the $\bar\Gamma_w=100$ evolution deviates less from
$\sigma_{flat}$ than the one for $\bar\Gamma_w=10$; this is because
the local disk-warp is smaller in the former case
($\partial\hatl/\partial\ln r\lesssim 0.24$ as compared with
$\partial\hatl/\partial\ln r\lesssim 0.94$), so that less additional
advection is introduced by the warping.  The strong dependence of
$\omega_p$ on $\bar\Gamma_w$ opens the possibility to account for a
wide variety of low-frequency quasi-periodic oscillations with the
nonlinear behavior of a warped disk driven by the magnetic warping
instability (see \S 5).

We now use eq.~(\ref{eq:D}) for the function $D(x)$
and extend the parameter searches to $\bar\Omega_p$,
$\cJ_{in}$ and $\eta_0$.  Figure~\ref{fig:other_params} presents
results for different choices of these parameters.  For each curve,
evolutions were first performed for large $\bar\Gamma_w$ until the steady
state is reached. Then $\bar\Gamma_w$ was gradually lowered so that a sequence
of steady-state solutions was traced out.  (When $\bar\Gamma_w$
becomes smaller than a certain critical torque, the disks suddenly
flattens completely; see \S 4.3.)

Several general features of the steady-state disks appear to be
robust: In all cases, $\omega_p$ strongly depends on $\bar\Gamma_w$.
Furthermore, $\beta_{in}$ becomes significant ($\gtrsim 60^\circ$) for
$\bar\Gamma_w$ even slightly exceeding the critical value for the
onset of the warping instability.  The precession frequency $\omega_p$
depends strongly on $\bar\Omega_p$ as well, as comparison between the
$\bar\Omega_p=10$ and $\bar\Omega_p=25$ curves reveals: For the same
$\bar\Gamma_w$, changing $\bar\Omega_p$ by a factor of 2.5 results in
a change of $\omega_p$ by about one order of magnitude.

\begin{figure}
\epsscale{0.6}
\plotone{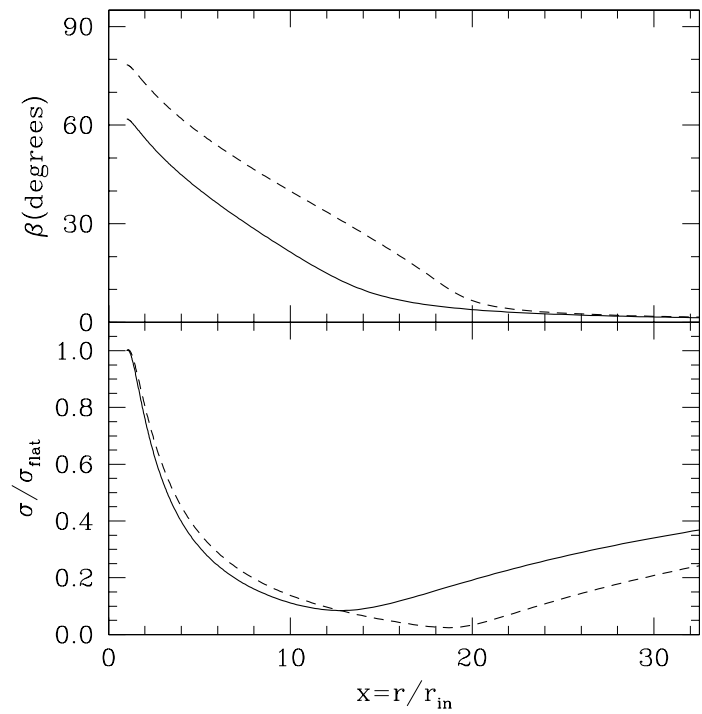}
\caption{\label{fig:compare_profiles_50}The
steady-state disk profiles for $\bar\Gamma_w=50$ (solid lines) and
$\bar\Gamma_w=100$, both with $\eta_0=50$, 
$D(x)$ from eq.~(\ref{eq:D}) and the other parameters as in
Fig.~\ref{fig:vary_GammaW}.  The top panel shows the tilt-angle
$\beta$ as a function of radius, the bottom panel the surface density
(cf. Fig.~\ref{fig:compare_profiles}.}
\end{figure}

Consider now the effect of larger viscosity ratio 
$\eta_0=\nu_{20}/\nu_{10}$ [$=12.5$ and 50, corresponding to
$\alpha=0.2$ and $0.1$ in $\nu_2/\nu_1=1/(2\alpha^2)$; see
\S 3.2]. A larger $\nu_2/\nu_1$ represents a disk more resistive to
bending, so that a larger warping torque $\bar\Gamma_w$ is required
for the magnetic warping instability to operate.  Also, $\beta_{in}$
approaches $90^\circ$ more slowly as $\bar\Gamma_w$ is increased.
Figure~\ref{fig:compare_profiles_50} shows the steady-state disk
profiles for the case of $\eta_0=50$.  These differ
appreciably from the profiles obtained with $\eta_0=1$ (as shown 
in Fig.~\ref{fig:compare_profiles}): The tilted region of the disk
extends to larger radii ($x\approx 20$) for $\eta_0=50$ 
than for $\eta_0=1$; the disk is also locally flatter 
($\partial\hatl/\partial\ln r\lesssim 0.13$) and
more depleted (with $\sigma/\sigma_{flat}$ being as small as $0.03$).

The inner boundary condition is somewhat critical. The disk behavior
described above works for relatively large $\cJ_{in}$ ($\go 0.2$;
depending on the other parameters). However, reducing $\cJ_{in}$
further leads to singularities in the disk: For evolutions with
smaller $\cJ_{in}$, say, $\cJ_{in}=0.1$, it appears that the surface
density tends to zero at some finite radius $x_{sing}$ away from the
inner edge. The tilt angle $\beta$ is non-zero inside $x_{sing}$ and
close to zero outside $x_{sing}$.  Overall, it appears that the disk
pinches off and two regions remain: a flat outer disk with inner
boundary at $x_{sing}$, and a strongly tilted, rapidly precessing
inner disk inside $x_{sing}$ (which may be viewed as part of the
magnetosphere).

\subsection{Discontinuous Behavior: Hystereses}
\label{sec:Hysteresis}

\begin{figure}
\epsscale{0.6}
\plotone{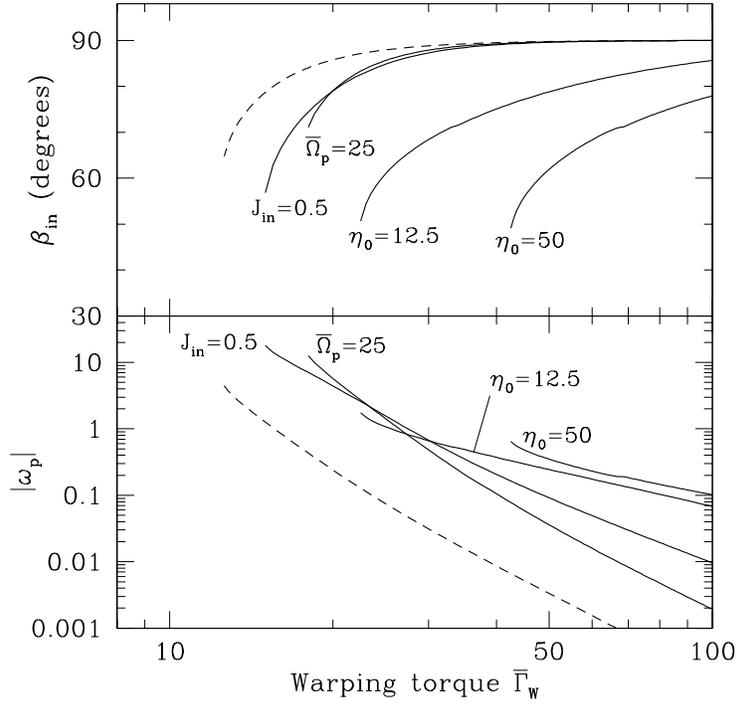}
\caption{\label{fig:other_params} Similar to
Fig.~\ref{fig:vary_GammaW}, but for different sets of parameters.  The
short dashed lines correspond to a run with the same parameters as in
Fig.~\ref{fig:vary_GammaW}, $\bar\Omega_p=10$, $\cJ_{in}=1$,
$\eta_0=1$, but with $D(x)$ given by Eq.~(\ref{eq:D}). Each one of the
remaining runs is obtained by changing the value of one of
these parameters, as labeled.}
\end{figure}

In Figure~\ref{fig:other_params}, the tilt angle
$\beta_{in}$ along each curve does not approach zero continuously as
$\bar\Gamma_w$ is reduced, but exhibits a discontinuity. A closer examination
reveals a hysteretic behavior of warp disks.

\begin{figure}
\epsscale{0.6}
\plotone{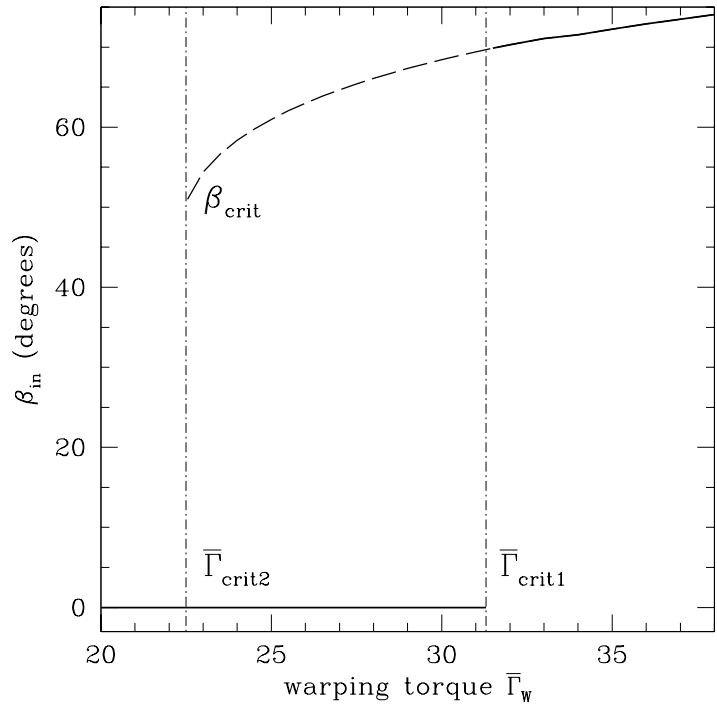}
\caption{\label{fig:instability_onset} Hystereses of warped disks.
Plotted is the steady-state tilt angle at the inner disk edge as a
function of the torque parameter $\bar\Gamma_w$.  See \S4.3 for
details. }
\end{figure}

Consider the $\eta_0=12.5$ case in Fig.~\ref{fig:other_params}
as an example (see Fig.~\ref{fig:instability_onset}).
If we start with a slightly perturbed disk ($\beta\ll 1$,
$\sigma=\sigma_{flat}$), then the perturbation decays for
$\bar\Gamma_w<\bar\Gamma_{crit\,1}\approx 31.3$ and the disk settles
down into the flat state.  For $\bar\Gamma_w$ slightly larger than
$\bar\Gamma_{crit\,1}$, the growth rate of the perturbation is
proportional to $\bar\Gamma_w-\bar\Gamma_{crit\,1}$;
nonetheless the perturbation grows to large amplitude (with
$\beta_{in}$ reaching $\approx 70^\circ$ in the final steady state).
Thus, starting from small perturbations, the disk evolves into a
steady state which depends discontinuously on 
$\bar\Gamma_w$ (see the solid line of Fig.~\ref{fig:instability_onset}).

The situation is different if one starts from a highly
warped state: We first evolve the disk with
$\bar\Gamma_w>\bar\Gamma_{crit\,1}$ until the final (strongly warped)
steady state is reached. We then reduce $\bar\Gamma_w$ gradually; for
each new value of $\bar\Gamma_w$ we evolve the disk to a new steady
state (using the ``previous'' steady-state solution as the initial
condition).  As long as $\bar\Gamma_w>\bar\Gamma_{crit,1}$ this
procedure produces the same sequence of steady-state disks we found
when starting from a slightly perturbed disk. However, we find that
even below $\bar\Gamma_{crit,1}$, there still exits steady-state
warped disks (see the dashed curve in
Fig.~\ref{fig:instability_onset}).  We can follow this sequence down
to a second critical value $\bar\Gamma_{crit,2}\approx 22.5$. Slightly
above $\bar\Gamma_{crit,2}$, the tilt angle depends on $\bar\Gamma_w$
as $\beta_{in}=\beta_{crit}+a \left(\bar\Gamma_w-
\bar\Gamma_{crit,2}\right)^{1/2}$ for some constant $a$ and
$\beta_{crit}\approx 50^\circ$.  As we reduce $\bar\Gamma_w$ below
$\bar\Gamma_{crit,2}$, this warped steady-state sequence terminates
abruptly and the disk relaxes to the flat state.

Thus, for $\bar\Gamma_{crit,2}<\bar\Gamma_w<\bar\Gamma_{crit,1}$, 
both a flat disk and a warped steady-state disk solution exist. 
The steady state for such a disk depends on the 
history of the disk. When the disk
parameters vary slightly --- for example because the accretion rate
varies --- the disk might switch discontinuously between the two
states. As the disk moves across a critical value, a strongly warped
disk might suddenly ``turn off'' and become flat, or a flat disk might
suddenly become unstable with the warp growing to significant values.
Since the local disk warp is fairly small at each radius (the maximum
$|\partial\hatl/\partial\ln r|$ being approximately $0.3$), it seems
unlikely that the hysteresis is merely an artifact of the adopted
simplifying evolution equations (see \S 3).

We find that all parameter sets examined in 
Fig.~\ref{fig:other_params} exhibit such a hysteresis
(e.g., for $\bar\Omega_p=10, \cJ_{in}=\eta_0=1$:
$\bar\Gamma_{crit,2}\approx 12.5, \bar\Gamma_{crit,1}\approx 16.9$;
for $\bar\Omega_p=25, \cJ_{in}=\eta_0=1$:
$\bar\Gamma_{crit,2}\approx 18, \bar\Gamma_{crit,1}\approx 38.2$;
for $\bar\Omega_p=10, \cJ_{in}=0.5, \eta_0=1$:
$\bar\Gamma_{crit,2}\approx 15, \bar\Gamma_{crit,1}\approx 20.3$;
for $\bar\Omega_p=10, \cJ_{in}=1, \eta_0=50$:
$\bar\Gamma_{crit,2}\approx 42.5, \bar\Gamma_{crit,1}\approx 69.7$).

\section{Discussion and Conclusion}

In this paper we have demonstrated through numerical simulations that
accretion disks around magnetic stars can develop into a warped 
steady-state in which the disk attains a fixed (warped) shape and precesses
rigidly.  This work extends our previous (local and global) linear
analysis of magnetically driven warping and precession of accretion
disks. We find that whenever the magnetic warping instability
criterion is satisfied in the linear regime, the inner disk (close to
the magnetosphere boundary) is likely to be highly warped. The
steady-state precession frequency spans a wide range, depending
sensitively on the warping and precession torques (and hence on the
parameters of the system).

\subsection{Limitations of the Models}

Before drawing any astrophysical conclusion of our work, it is useful
to recall some of the limitations of our models. First, the
nonlinear warp equations we adopted in our simulations are based on
phenomenological descriptions of viscosities (see \S 3).
Second, the magnetic torques formulae were derived under several
assumptions/ansatz: e.g., locally flat disk, the use of the free parameter 
$\zeta$ to parametrize the magnetic field twist in quasi-steady state
(see \S 3). Finally, there are intrinsic uncertainties associated with 
the inner boundary conditions of the disk. In our problem, the inner 
boundary is located at the magnetosphere radius, where the transition between a
Keplerian disk and a corotating magnetosphere occurs. The physics that
determines this transition is obviously complicated (see references
cited in \S 1). The magnetic warping and precession torques are steep
functions of radius and are maximal close to the inner disk
edge. Therefore the details of our numerical results depend
sensitively on the physics at the inner radius of the accretion disk.
In our calculations we adopted the simplest inner boundary conditions
for the disk. A more rigorous resolution of this problem will have to
involve studying the coupling of the warped disk and the
magnetosphere.

Given these uncertainties and limitations, we cannot be sure whether 
some of the features we found for the warped disks (such the 
hysteresis behavior discussed in \S4.3) correspond to reality in any way
or simply represent artifacts of our models. Nevertheless,
our numerical experiments show that the general behaviors of 
magnetically driven warped disks that we found (as summarized at
the beginning of \S5) are robust.

\subsection{Astrophysical Implications}

The most important feature of magnetically driven disk warping 
studied in this paper is that the disk naturally evolves into 
a steady state which is highly warped near the 
inner radius (the magnetosphere boundary). This is in contrast to other 
warping mechanisms (e.g., those due to radiation pressure or 
irradiation-driven disk wind) which operate from outside-in.
Such a highly warped disk can lead to modulation of the 
observed radiation flux either by obscuring the central star or
by changing the reprocessed disk emission. As mentioned in \S 1,
there is growing observational evidence for quasi-periodic oscillations 
(QPOs) or variabilities in radiation fluxes induced by warped inner disks 
in various systems, ranging from YSOs (e.g. AA Tauri; see Bouvier 
et al.~1999) to X-ray binaries (e.g. Her X-1: see Deeter et al.~1998;
Scott et al.~2000; see also Shirakawa \& Lai 2002b for a list of other 
X-ray binaries showing mHz QPOs).

Another interesting feature of our models is that the disk can be either 
in a completely flat state (warping stable) or in a highly warped 
state. Thus for a given system, as the disk parameters (e.g. accretion
rate) vary, the disk can switch abruptly between the two states, leading to 
the appearance/disappearance of QPOs (with the corresponding change
in radiation flux or spectrum). Indeed, in X-ray binaries, there are
many examples where QPOs occur only in certain spectral states and 
not in the others (e.g., van der Klis 2000).

Our calculations showed that the steady-state precession frequency of
the disk can be much smaller (by up to several orders of magnitude,
depending on the system parameters, cf. Fig.~\ref{fig:other_params})
than the frequency of the linear mode; the latter was approximately
equal to the precession frequency at the disk inner radius (see
Shirakawa \& Lai 2002a,b for typical numbers for X-ray binaries).
Thus, for magnetized neutron stars (with surface magnetic field $\sim
10^{12}$~G), the global precession frequency is
\begin{equation}
 \nu_p=A\frac{\Omega_p(r_{in})}{2\pi}=-(1.2~{\rm
mHz})\, A\,\mu_{30}^2M_{1.4}^{-1/2}r_8^{-11/2}\Sigma_4^{-1}
D^{-1}\sin^2\theta, 
\end{equation}
where the dimensionless parameter $A$ can be much smaller than unity, and
$\mu_{30}=\mu/(10^{30}~{\rm G}\,{\rm cm}^3)$,
$M_{1.4}=M/(1.4M_{\odot})$, $\Sigma_4=\Sigma(r_{in})/(10^4~{\rm
g\,cm}^{-2})$, with $r_8=r_{in}/(10^8~{\rm cm})=
3.4\,\eta\,\mu_{30}^{4/7}M_{1.4}^{-1/7}{\dot M}_{17}^{-2/7}$ as well as
${\dot M_{17}}=\dot M/(10^{17}\,{\rm g\,s^{-1}})$. For typical
parameters appropriate for T Tauri stars, the precession period is
given by
\begin{equation}
\nu_p^{-1}=\frac{2\pi}{A\Omega_p(r_{in})}=-(124\,{\rm
yrs}) A^{-1}B_3^{-2}\,R_2^{-6}M_1^{1/2} \!\left(\frac{r_{\rm
in}}{8R_\odot}\right)^{11/2}\!\!  \Sigma_3\,D\,(\sin\theta)^{-2},
\end{equation}
where $B=\mu/R^3=10^3B_3$~G is the surface magnetic field of the star,
$R=(2R_\odot) R_2$ is the stellar radius, and $M_1=M/(1~M_\odot)$.
Therefore magnetically driven precession can potentially explain some
long-period QPOs/variabilities in X-ray binaries and YSOs\footnote{ We
have only included the magnetic torques in our calculations. In real
astrophysical systems, there may be other torques that contribute to
(or even dominate) the disk precession; for example the tidal torque
in binary systems (e.g. Terquem et al.~1999; Ogilvie \& Dubus 2001)
and the torque associated with general relativistic frame dragging
effect in low-mass X-ray binaries (see Shirakawa \& Lai 2002a).  Note
that even in those systems where other torques dominate disk
precession, the magnetic warping torque studied here can still be
important in exciting/maintaining the disk tilt.}. It is of interest
to note that the magnetically driven precession is retrograde, as
observed in some systems (e.g., Her X-1).

Finally, if the outflows from YSOs are produced from interaction
between the stellar magnetic field and the disk, as in the X-wind
models (Shu et al.~1994), then the magnetic effects studied in this
paper may be responsible for the jet precession observed in several
systems [see Terquem et al.~1999 and references therein; see also Lai
(2003) for possible warping instability in magnetically driven disk
outflows].

\acknowledgments This work has been supported in part by NSF Grants
AST-9986740, AST-0307252, PHY-9900672 and NASA grant NAG 5-12034.

\appendix
\section{Numerical method}

Equations~(\ref{eq:sigma2}) and~(\ref{eq:lhat2}) are essentially of the form
\begin{align}\label{eq:sigma3}
\dnachd{\sigma}{\tau}-A\dnachd{^2\sigma}{x^2}-B\dnachd{\sigma}{x}&=C,\\
\label{eq:hatl3}
\dnachd{\ve l}{\tau}-D\dnachd{^2\ve l}{x^2}-E\dnachd{\ve l}{x}&=F,
\end{align}
with coefficients $A,\ldots, F$ depending nonlinearly on the variables
$\sigma$ and $\ve l$ (for ease of notation, we omit the hat on
$\hatl$ here).  We discretize these equations using the
Crank-Nicholson method (Press et al.~1992).  Time-derivatives are
discretized as usual, for example
\begin{equation}
\left(\dnachd{\sigma}{\tau}\right)_i\to
\frac{\bar\sigma_i-\sigma_i}{\Delta\tau}.
\end{equation}
The index $i$ labels the spatial grid-points, unbarred quantities like
$\sigma_i$ denote values at the current time $\tau_0$ (which are known),
and barred quantities denote values at the new time
$\tau_0+\Delta\tau$ which are to be determined.  Spatial derivatives
are averaged over $\tau_0$ and $\tau_0+\Delta\tau$, for example
\begin{equation}\label{eq:spatial-difference1}
\left(A\dnachd{^2\sigma}{x^2}\right)_i
\to
-\frac{A_i}{2}\frac{\sigma_{i+1}-2\sigma_i+\sigma_{i-1}}{(\Delta x)^2}
-\frac{\bar A_i}{2}\frac{\bar\sigma_{i+1}-2\bar\sigma_i+\bar\sigma_{i-1}}
{(\Delta x)^2}.
\end{equation}
Here, $A_i$ is computed using the known values at $\tau_0$, whereas
$\bar A_i$ is a function of the desired values at $\tau_0+\Delta\tau$.
This discretization leads to a nonlinear system of equations, which is
solved iteratively: We start with a guess for $\sigma$ and $\ve l$ at time
$\tau_0+\Delta\tau$, which we denote with tildes, $\tilde\sigma_i,
\tilde{\ve l}_i$.  Based on these guesses, we compute coefficients
$\tilde A_i, \ldots \tilde F_i$.  Now, we discretize spatial
derivatives like (\ref{eq:spatial-difference1}) using $\tilde A_i$
instead of $\bar A_i$:
\begin{equation}\label{eq:spatial-difference2}
\left(A\dnachd{^2\sigma}{x^2}\right)_i
\to
-\frac{A_i}{2}\frac{\sigma_{i+1}-2\sigma_i+\sigma_{i-1}}{(\Delta x)^2}
-\frac{\tilde A_i}{2}\frac{\bar\sigma_{i+1}-2\bar\sigma_i+\bar\sigma_{i-1}}
{(\Delta x)^2}.
\end{equation}
Substituting these difference expressions into Eqs.~(\ref{eq:sigma3})
and~(\ref{eq:hatl3}) now yields {\em linear} tridiagonal systems of
equations for $\bar\sigma_i$ and $\bar{\ve l}_i$.  Solving these
yields improved values $(\bar\sigma_i, \bar{\ve l}_i)$ for the
variables at the next time step which are used in place of
$(\tilde\sigma_i, \tilde{\ve l}_i)$.  We iterate this process until
convergence.

This scheme is computationally more expensive per time-step than an
explicit method.  However, it is second order accurate in time and
unconditionally stable, so that one can take very large time-steps.
In practice, we often exceed Courant-factors of 1000.  These large
time-steps are especially important during the slow relaxation toward
a coherently rotating steady-state disk.  In such a case, one is
primarily interested in the time-independent final state, whereas the
transient evolution toward this state is less important.

The code uses an adaptive time step: $\Delta\tau$ is adjusted such
that the difference in solution at $\tau_0$ and $\tau_0+\Delta\tau$ is
smaller than some threshold, typically $10^{-4}$.  Furthermore, the
code can be used with different radial distributions of grid points,
for example linearly or logarithmically spaced grid points.  This is
done via a mapping $\bar x\to x=x(\bar x)$, which relates a
``computational'' coordinate $\bar x$ (in which the grid is uniform)
to the ``physical'' radial coordinate $x$.  Rewriting ``physical''
derivatives in terms of the computational coordinate $\bar x$,
$$
\dnachd{}{x}=\dnachd{\bar x}{x}\dnachd{}{\bar x},
\qquad\qquad
\dnachd{^2}{x^2}=\left(\dnachd{\bar x}{x}\right)^2\dnachd{^2}{\bar x^2}
+\dnachd{^2\bar x}{x^2}\dnachd{}{\bar x},
$$ shows that the coefficients $A, \ldots F$ will be modified by this
mapping, however, the principal structure of Eqs.~(\ref{eq:sigma3})
and (\ref{eq:hatl3}) does not change.  Usually, logarithmically spaced
grid-points are used which accurately resolve the fine structure close
to the inner edge, while allowing to move the outer boundary far out,
typically to $x_{out}=1000$.

We verified that our code is second order convergent in space and time
and recovers the standard flat disk solution,
Eq.~(\ref{eq:Sigma-flat}), in the absence of warping/precession
torques.  Section \ref{sec:LinearEvolution} confirms that in the
linear regime our code recovers the linear stability analysis of
Shirakawa \& Lai (2002b).


\clearpage

\end{document}